\documentstyle[psfig,epsf]{mn2e}

\newcommand{\ha}{H$\alpha$} 
\newcommand{\hbeta}{H$\beta$}
\newcommand{\helium}{He{\sc i}}
\newcommand{\heliumb}{He{\sc ii}}

\newcommand{\NII}{[N~{\sc ii}]~6548~\&~6584~\AA}

\newcommand{\oiii}{[O~{\sc iii}]~5007~\AA}

\newcommand{\oii}{[O~{\sc ii}]} 

\newcommand{\nitrogen}{[N~{\sc ii}]}
\newcommand{\oxygen}{[O~{\sc iii}]}

\newcommand{\sulfurt}{[S~{\sc ii}]}
\newcommand{\flux}{$10^{-16}$ erg s$^{-1}$ cm$^{-2}$ arcsec$^{-2}$}
\title[New Planetary Nebulae in the Galactic Bulge region with
$l>0^{o}$] {New Planetary Nebulae in the Galactic Bulge region with
$l>0^{o}$~- I. Discovery method and first results}

\author[P. Boumis et al.] {P. Boumis$^{1}\thanks{e-mail:
ptb@astro.noa.gr}\thanks{Present address: Institute of Astronomy \& Astrophysics,
National Observatory of Athens, I. Metaxa \& V. Paulou, GR--152 36 P. Penteli,
Athens, Greece}$, E. V. Paleologou$^{1}$,
F. Mavromatakis$^{1}$ and J. Papamastorakis$^{1,2}$\\ $^{1}$Department
of Physics,University of Crete, P.O. Box 2208, GR-710 03 Heraklion,
Crete, Greece.\\$^{2}$Foundation for Research and Technology-Hellas,
P.O. Box 1527, GR-711 10 Heraklion, Crete, Greece.\\}

\date{Received 2002 April 5. Accepted 2002 November 1}
\pagerange{\pageref{firstpage}--\pageref{lastpage}}

\begin{document}  

\maketitle

\begin{abstract}
 
\noindent We present the first results of an \oiii\ interference
filter survey for Planetary Nebulae (PNe) in the Galactic
bulge. Covering (at first) the 66 per cent of the survey area, we
detected a total of 90 objects, including 25 new PNe, 57 known PNe and
8 known PNe candidates. Deep \ha $+$\nitrogen~CCD images have been
obtained as well as low resolution spectra for the newly discovered
PNe. Their spectral signature suggests that the detected emission
originates from a photoionized nebula. In addition, absolute line
fluxes have been measured and the electron densities are
given. Accurate optical positions and optical diameters were also
determined.
\end{abstract}


\section{Introduction}

The Galactic Planetary Nebulae are of great interest because of their
important role to the chemical enrichment history of the interstellar
medium as well as in the star formation history and evolution of our
Galaxy (Beaulieu et al. 2000 and references therein). The study of PNe
offers the opportunity to determine basic physical parameters (like
morphology, kinematics, abundances, distances, masses etc.), which
will help to test theoretical models and understand more about the
bulge dynamics (see sect. 2). The distance determinations of PNe in
the Galactic bulge suggest distances at $\sim$8 kpc (7.8 kpc; Feast
1987, 8.3 $\pm$ 2.6 kpc; Schneider \& Buckley 1996). The Galactic
bulge extends over a region of galactic longitude $l \simeq \pm
15^{o}$ and galactic latitude $b
\simeq \pm 10^{o}$ (dimensions $\sim$2 kpc toward and $\sim$1.5 kpc
perpendicular the Galactic plane, respectively - Weiland et
al. 1993). In total, the bulge contains about 10$^{11}$ stars of
Population II as well as gas ($\sim 10^{6} {\rm M}_{\odot}$) and dust
(mass $\sim$100 times less than the mass of the gas - Morison \&
Harding 1993). It is generally accepted that a large number of PNe
exists in the Galactic bulge, but because of the excessive
interstellar extinction, only a small number of them (less than 500)
have been discovered (K\"{o}ppen \& Vergely 1998; Parker et
al. 2001b). According to Acker et al. (1992a) and Zijlstra \& Pottasch
(1991), the total number of PNe in the Galaxy is estimated between
15,000 to 30,000. Approximately, 3,400 PNe have been discovered up to
date, meaning that only a small number of all PNe have been
detected. In particular, Acker et al. (1992b, 1996) report 2204
objects which have been classified as true, probable and possible
PNe. Most of them were found through surveys in different wavebands
(optical, radio, infrared - Moreno et al. 1988; Pottasch et al. 1988;
Ratag 1990; van de Steene 1995, etc.). Kohoutek (2001) presented 169
additional new PNe which were discovered through different surveys
between the years 1995-1999, while Cappellaro et al. (2001) published
a catalogue of 16 new PNe discovered towards the galactic center. In a
very recent catalogue, Parker et al. (2001b) presented more than 900
new and candidate PNe which were discovered during the AAO/UKST \ha\
survey of PNe in the Galactic plane. Note that not all surveys
utilized narrow band imagery to identify candidate PNe (e.g. Lauberts
1982).

The task of determining the wavelength band suitable for a search of
new PNe in the bulges is under debate. The interstellar extinction
being higher in the direction of the bulge, reduces the effectiveness
of the optical surveys.  For this reason, radio continuum and
far-infrared surveys were made in regions with high extinction,
resulting to new PNe discoveries. On the other hand, as Beaulieu,
Dopita \& Freeman (1999) noted long waveband surveys work well only
with dusty (young) PNe while the low-density (old) PNe are
undetectable by IRAS. Furthermore, it is well known that most PNe emit
strongly in the \ha~and \oiii~lines, and that the latter is
affected more by interstellar extinction than \ha. Moreover, [O~{\sc iii}]
bright PNe display \ha/\oiii\ $<$ 1 and mainly the low-density
PNe. Recent high resolution spectrophotometric observations (Cuisinier et
al. 2000; Escudero \& Costa 2001) determined the abundances of known
PNe in the Galactic bulge. Using these results, a comparison between
the \ha~and [O~{\sc iii}] line intensities shows that 40 per cent of
them display \ha/\oiii $<$ 1.

The right combination of the telescope size and the CCD camera,
depending on the distance to search, can provide a useful tool for the
detection of new PNe (Ciardullo et al. 1989; Jacoby et al. 1990;
Feldmeier, Ciardullo \& Jacoby 1997 and references therein).
Therefore, we decided to performed survey observations of the Galactic
bulge using the 0.3 m telescope of a wide field of view (combined with
a highly efficient CCD camera) and a narrow \oiii~filter. The \oiii\
filter (like any other emission line filter) may not be the best
choice for the detection of new PNe but note that a significant number
of PNe are characterized by strong \oiii\ emission relative to \ha.
We must note that a PN discovery has already been done in the past
using this configuration (Xilouris et al. 1994).

Informations concerning the importance of PNe in the bulge are given
is sect. 2, the \oxygen~survey (observations and analysis) are given
in sect. 3, while in sect. 4 we present follow-up observations (CCD
imaging and spectroscopy) of the newly discovered PNe. In addition,
flux measurements, diameter determination, accurate positions and
other physical properties for the new PNe are given is sect. 5. We
must note that this paper is the first of a sequence, where we will
present in detail imaging and spectrophotometric results for all the
new PNe discovered in our survey, as well as fitted models to describe
their physical parameters. In particular, images and spectra for the
new PNe found in the remaining fields will be presented, while flux
calibrated images for most of the new PNe will be used in PNe models
(e.g. van Hoof \& van de Steene 1999) in order to determine their
physical parameters (like abundances, distances etc).

\section{Planetary nebulae as tracers of the Galactic bulge}

Many tracers have been used in the past to study the
Galactic bulge. In particular, the OH/IR stars (e.g. Sevenster 1999),
the M giant stars (e.g. Minniti et al. 1996a; Minniti 1996b), the
Miras and other LPVs (Whitelock 1994), the carbon stars (Whitelock
1993), the K giants stars (e.g. Minniti et al. 1995; Minniti 1996b),
the RR Lyrae stars (Walker \& Terndrup 1991) and the PNe (Kinman,
Feast \& Lasker 1988; Durand, Acker \& Zilstra 1998; Beaulieu et
al. 2000). The first three (OH/IR, Miras and M giant stars) are biased
toward the metal--rich population. The carbon stars are also an
indication of metal--rich population but they are not very numerous in
the Galaxy. The K giant stars have been used a lot as tracers since
they can be found at all metallicities but they are biased because of
their faintness. The RR Lyrae stars represent the metal--poor tail of
the bulge metallicity distribution and they are also faint.

On the other hand, PNe are not biased toward the metal--rich
population (Hui et al. 1993). They originate from intermediate and low
initial mass stars and therefore constitute a relatively old
population. The He and N abundances are modified along the stellar
evolution of the progenitor and can be related to their masses and
age. Their optical spectrum is dominated by strong emission lines
(like \ha\ and \oxygen) allowing accurate velocity measurements. Up to
now, their distances are not well defined since it is difficult to
determine their absolute diameter which is necessary for distance
calculation (e.g. Schneider \& Buckley 1996). However, Gathier et
al. (1983) estimated that a large number of PNe with small diameters
($\leq$ 20\arcsec) belong to the Galactic bulge according their
angular diameter, spatial distribution and radial velocities
measurements. Taking into account all the parameters discussed in this
section, the Galactic bulge PNe appear as a very important tool in
order to study the dynamics and kinematics of the Galactic bulge.

\section{The \oiii\ Survey}

\subsection{Observations}

The observations reported here were made with the 0.3 m
Schmidt-Cassegrain (f/3.2) telescope at Skinakas Observatory in Crete,
Greece. An \oiii~interference filter with an 28\AA~bandwidth was used
in combination with a Thomson CCD (1024$\times$1024, 19$\times$19
$\mu$m$^{2}$ pixels). This configuration results in a scale of
4\arcsec.12 pixel$^{-1}$~and a field of view of
71\arcmin~$\times$~71\arcmin on the sky. In our survey we observed the
regions $10^{o} < l < 20^{o}$, $-10^{o} < b < -3^{o}$ and $0^{o} < l <
20^{o}$, $3^{o} < b < 10^{o}$ (named Field A and Field B, respectively
in Fig. 1). The reasons to do this was (a) the site of the Observatory
which allows us to observe down to $-25^{o}$ declination and (b) the
\oiii~emission line which is heavily
absorbed between $-3^{o} < b < 3^{o}$. In our survey, we tend to
discover planetary nebulae which (1) characterized by strong \oiii\
emission, (2) are slightly extended because of our spatial resolution
(3) are point--like with signal to noise greater than 4 (4) do not
have declination below -25$^{o}$ and (5) are observed only during dark
time periods to avoid the moonlight scattering effect to the
\oiii~line. We must note that the area of the Galactic Bulge was also
covered by the AAO/UKST \ha\ survey (Parker et al. 2001a -- in fact
they have covered the total area of the southern Galactic plane). In
Fig. 1 the filled rectangles represent the observed fields. We covered
at first a region of 66 per cent of the proposed grid (116 fields out
of 179) because of the availability of the telescope with respect to
the time needed to complete the survey. . The observational details
are given in Table~\ref{table1}.  During the 2001 observing season we
observed the remaining 62 fields (Boumis et al. 2003 - Paper II). All
the objects were observed between airmass 1.4 to 2.0 in similar
observing and seeing conditions.

Two exposures in \oiii~of 1200 s and three exposures in the continuum
of 180 s were taken to make sure that any cosmic ray hits will be
identified and removed successfully (see section 3.2 for the image
processing details). Two different continuum filters were used,
dependent on their availability. Details about the filters are given
in Table~\ref{table2}.

\begin{table}
\centering
\caption{Observational details for the \oiii~Imaging Survey}
\label{table1}
\begin{tabular}{l|c|c}
\hline
Date &
Number of &
Number of \\
& observing nights & clear nights \\
\hline
2000 May 6-7      & 2 & 2 \\
2000 May 25-31    & 7 & 3 \\
2000 Jun 25-29    & 5 & 5 \\
2000 Jul 1-7      & 7 & 7 \\
2000 Jul 24-27    & 4 & 4 \\
2000 Jul 31-Aug 3 & 5 & 5 \\
2000 Aug 24-Aug 28 & 5 & 5 \\
2000 Aug 30-Sep 04 & 6 & 6 \\
\hline
\end{tabular}
\end{table}

\begin{table*}
\centering
\caption{Filters used for the Imaging Survey}
\label{table2}
\begin{tabular}{c|c|c|c|c}
\hline
Filter &
Central wavelength &
FWHM &
Type &
Manufacturer \\
 & ($\AA$) & ($\AA$) & & \\
\hline
\oiii  & 5005 & 28 & Interference & Spectrofilm \\
Continuum & 5470 & 230 & Str\"{o}mgren--y & Custom Scientific \\
Continuum & 6096 & 134 & Interference & Spectrofilm \\
\hline
\end{tabular}
\end{table*}

\subsection{Analysis}

Standard IRAF, MIDAS and STARLINK routines were used for the reduction
of the data. The analysis was performed by using three different
packages since different tasks were easier and more accurate to
perform in certain packages than in others. Individual images were
bias subtracted and flat--field corrected through observations of the
twilight sky. In particular, approximately 15 bias frames and 11
twilight flat--field frames (in each filter) were obtained each night
and the master bias and flat--field frames (for each filter) were used
in the image processing. The sky background has also been subtracted
from each individual frame. No dark frames were used, since the dark
current was negligible for the exposure time used.

\subsubsection{Selection of the PNe candidates}

It is difficult to detect faint emission sources in a very dense star
field as in the Galactic bulge. Therefore, special techniques must be
followed to attack this problem. It is well known that for the
detection of faint emission line sources - like the PNe in our case -
a narrow-band filter (i.e. [O{\sc iii}]) should be used in combination
with a wider-band filter (continuum) where no significant emission
lines exist and only the field stars appear. The correct scaling and
subtraction of the two images will suppress the field stars and only
the [O{\sc iii}] emitters and possible cosmic rays will still be
visible.
 
Our detection method (Boumis \& Papamastorakis 2002) is similar but
not the same to that followed in the Galactic bulge by Beaulieu et
al. (1999). Their idea was to combine two \ha~frames and then divide
the summed image by the continuum image for each field. Here, instead
of division, we identify the PNe candidates by following another
approach. In particular, after calculating the correct scaling between
the stars from the [O{\sc iii}] and the continuum images (the three
continuum images of each field were combined to produce a master
continuum), we subtracted from each of the two [O{\sc iii}] images the
corresponding master continuum image for each field. In order to do
the correct scaling (match the total star signal above the sky in both
images), we chose a box of a specific number of pixels where non
saturated stars were included. The continuum images were multiplied by
a factor f (similar to that given by Beaulieu et al. 1999) and then
their total star signal was the same as for the [O{\sc iii}]
images. The initial PNe identification was performed by ``blinking''
each continuum subtracted [O{\sc iii}] frame and the corresponding
continuum. Given the angular size of $\sim$4 arcsec of a CCD pixel,
the PNe candidates at the bulge appeared as slightly extended (or
sometimes point--like) sources on the [O{\sc iii}] images, but were
absent in the continuum. As mentioned above, the new sources could be
PNe candidates or cosmic rays. To make sure that no cosmic rays were
included among the candidates, we checked visually each individual
[O{\sc iii}] frame for each field. Objects that appeared bright in one
frame but were absent in the other, were considered as cosmic ray hits
because a real emission source should be present in both [O{\sc iii}]
images.  Another limitation was that all point--like objects have a
signal to noise greater than 4 according to their count measurements
since the survey images are not flux calibrated. We must note that we
have tried to follow Beaulieu et al. (1999) method and the two [O{\sc
iii}] images of each field were combined before subtraction in order
to increase the signal to noise ratio. However, the problem of
removing correctly the cosmic rays still remained and we decided to
perform the blinking technique between the two continuum subtracted
[O{\sc iii}] images and the corresponding master continuum image.
After a very detailed and systematic visual investigation of 116
fields, we identified 90 objects. The procedure of how the PNe
candidates looked like through the different steps can be seen in
Fig.~2, where there are examples of images (20\arcmin\ on both sides
each) for three different objects which show our survey's typical
discoveries. The first set of images (a) is a known PN which was found
and presented for comparison reasons; the other two sets (b) and (c)
are new discoveries. Note that the remaining black dots in the last
image of each set (except the PNe) are cosmic rays which were not
removed by the continuum subtraction but clearly identified since they
do not appear in both the continuum subtracted \oxygen\ images.

\subsubsection{Preliminary Astrometry \& Survey Results}

After identifying the PNe candidates, the next step was to perform an
astrometric solution for all images containing one or more candidates.
In order to calculate the equatorial (and galactic) coordinates of the
PNe candidates, we used the Hubble Space Telescope (HST) Guide Star
Catalog (Lasker, Russel \& Jenkner 1999) and IRAF routines (Image
package). Typical rms error in each frame was found to be
$\sim$1\arcsec.3 using the task ``ccmap'' in the Image package of
IRAF. The 90 objects were then checked in order to identify the
already known PNe. For that purpose we used any up--to--date published
catalogue related to planetary nebulae. A list of all catalogues used
is given in Table~\ref{catalogues}. The search showed that from the 90
objects found in our survey, the 57 are known PNe, 8 are known PNe
candidates and 25 are new PNe candidates. Note that independently from
our search, Parker et al. (2001b) presented a new catalogue of GBPNe
where 17 of our new PNe are included as new and candidate PNe (see
Table 4). In order to make sure that all catalogued PNe which fall in
our fields were detected, we performed a second search and found that
16 known PNe and 14 known PNe candidates are not included in our
detections. However, knowing their position a re--examination showed
that most of the ``missed'' known PNe and PNe candidates do appear in
our survey frames as very faint point sources due to their small
angular size and their very low signal to noise ratio ($\sim
1\sigma$). The 8 known PNe candidates which have been confirmed in our
survey (in galactic coordinates) are: 003.9+02.7, 004.4+06.4,
006.7+03.4, 007.8-03.8, 008.3+03.7, 009.3-06.5, 019.0-04.2 and
019.7-08.2. Note, that all the above PNe and PN candidates are listed in the
catalogues given in Table~\ref{catalogues}.

\begin{table*}
\centering
\caption{A list with all the planetary nebulae catalogues used. Note
that the Strasbourg-ESO Catalogue was used as a reference, because it
was the most complete catalogue up to 1992. Ratag's thesis was the
only exception as it was published before 1992 but it was not included in the
above catalogue.}
\label{catalogues}
\begin{tabular}{c|c|c}
\hline
N$_{o}$ & Catalogue & Reference\\
\hline
1 & A Study of GBPNe & Ratag 1990 \\
2 & Strasbourg-ESO Catalogue of GPNe (Part I, II) & Acker et al. 1992 \\
3 & Obscured PNe & van de Steene 1995 \\
4 & First Supplement to the Strasbourg-ESO Catalogue of GPNe & Acker
et al. 1996 \\
5 & IAC Morphological Catalog of Northern GPNe & Manchado et al. 1996 \\
6 & Catalogue of GPNe (1990-1994) & Kohoutek 1997 \\
7 & Innsbruck Data Base of GPNe & Kimeswenger, Kienel \& Widauer 1997 \\
8 & Kinematics of 867 GPNe & Durand, Acker \& Zijlstra 1998 \\
9 & Infrared PN in the NRAO VLA Sky Survey & Condon, Kaplan \& Terzian 1999 \\
10 & A Survey of PNe in the Southern Galactic Bulge & Beaulieu et al. 1999 \\
11 & Version 2000 of the Catalogue of GPNe & Kohoutek 2001 \\
12 & Radio Observations of New GBPNe & van de Steene et al. 2001 \\
13 & Optical Coordinates of Southern PNe & Kimeswenger 2001 \\
14 & New PNe towards the galactic center & Cappellaro et al. 2001 \\
15 & The Edinburgh/AAO/Strasbourg Catalogue of GPNe & Parker et al. 2001b \\
\hline
\end{tabular}
\end{table*}

\begin{table*}
\centering
\caption{Newly discovered PNe.}
\label{table5}
\begin{tabular}{c|c|c|c|c|c|c|c|c|c|c}
\hline
Object & PN G & RA & Dec & IRAS source  & F$_{12}$ & F$_{25}$ & F$_{60}$ & F$_{100}$ & F$_{\rm Quality}$ & Ref$^{\rm a}$ \\
 & (lll.l $\pm$ bb.b) & (J2000) & (J2000) & & (Jy) & (Jy) & (Jy) & (Jy) & (12,25,60,100 $\mu$m) & \\
\hline
PTB1  & 003.5$+$02.7 & 17 43 39.2 & $-$24 31 53.0 & 17405$-$2430 & 3.35 & 2.88 & 2.60 & 18.2 & 1,1,3,1 & 1a\\
PTB2  & 004.1$+$07.8 & 17 26 12.2 & $-$21 17 53.0 & & & & & & & 1a\\
PTB3  & 004.6$+$06.1 & 17 33 33.0 & $-$21 51 23.2 & & & & & & & 1b\\
PTB4  & 005.4$+$04.0 & 17 42 54.7 & $-$22 14 16.3 & & & & & & & 1a\\
PTB5  & 005.7$+$04.5 & 17 41 38.7 & $-$21 44 32.2 & 17386$-$2143 & 2.56 & 0.876 & 1.08 & 10.7 & 1,1,3,1 & \\
PTB6  & 006.1$+$03.9 & 17 45 06.5 & $-$21 41 55.1 & & & & & & & 1a\\
PTB7  & 007.1$+$07.4 & 17 34 30.2 & $-$19 01 01.1 & & & & & & & 1b\\
PTB8 & 007.2$+$03.4 & 17 49 13.7 & $-$21 01 42.9 & 17462$-$2100 & 1.19 & 0.712 & 1.41 & 12.1 & 1,3,3,1 & 1a\\
PTB9 & 007.2$+$04.9 & 17 43 34.0 & $-$20 13 55.7 & & & & & & & 1b\\
PTB10 & 007.3$+$03.5 & 17 49 26.7 & $-$20 54 31.1 & & & & & & \\
PTB11 & 007.4$-$03.0 & 18 13 40.2 & $-$23 57 38.5 & & & & & & \\
PTB12 & 007.9$+$04.3 & 17 47 15.6 & $-$19 57 27.6 & & & & & & & 1b\\
PTB13 & 008.8$+$03.8 & 17 51 08.7 & $-$19 25 46.6 & & & & & & & 2\\
PTB14 & 009.4$+$03.9 & 17 52 17.8 & $-$18 52 02.4 & & & & & & \\
PTB15 & 011.5$+$03.7 & 17 57 05.6 & $-$17 11 09.7 & 17541$-$1710 & 0.305 & 0.78 & 1.28 & 13.1 & 1,3,3,1 & 1a\\
PTB16 & 012.0$+$07.4 & 17 45 07.5 & $-$14 53 23.4 & & & & & & & 1a\\
PTB17 & 012.5$+$04.3 & 17 57 10.5 & $-$15 56 17.3 & & & & & & & 1a,2\\
PTB18 & 013.4$+$08.8 & 17 42 58.0 & $-$12 56 46.9 & & & & & & & 1a\\
PTB19 & 014.0$+$04.8 & 17 58 26.0 & $-$14 25 25.1 & 17555$-$1425 & 0.29 & 0.502 & 1.58 & 7.68 & 1,3,3,1 & \\
PTB20 & 016.1$+$07.7 & 17 52 15.0 & $-$11 10 35.6 & 17494$-$1109 & 0.269 & 0.466 & 0.71 & 3.49 & 1,3,3,1 & \\
PTB21 & 016.6$+$07.0 & 17 55 53.3 & $-$11 05 41.7 & 17529$-$1103 & 1.03 & 0.459 & 0.529 & 4.16 & 3,1,1,1 & \\
PTB22 & 016.8$+$07.0 & 17 56 21.4 & $-$10 57 34.3 & & & & & & & 1a\\
PTB23 & 018.0$-$02.2 & 18 31 50.6 & $-$14 15 27.5 & 18290$-$1417 & 1.42 & 1.77 & 17.2 & 182.0 & 1,3,1,1 & 1a\\
PTB24 & 018.4$+$05.3 & 18 05 28.2 & $-$10 20 42.6 & & & & & & & 1b\\
PTB25 & 018.8$-$01.9 & 18 32 04.6 & $-$13 26 12.7 & 18292$-$1328 & 1.97 & 1.14 & 22.1 & 301.0 & 1,3,1,1 & 1a\\
\hline
\end{tabular}

\medskip{}
\begin{flushleft}

${\rm ^a}$ Independently discovered by Parker et al. 2001b as new PNe (1a) and candidates PNe (1b), and Cappellaro et al. 2001 (2).\\  

\end{flushleft}
\end{table*}

\subsubsection{Search in other wavelengths}

We performed a search in the IRAS Point Source Catalog (1988) in order
to check for possible presence of dust to the new PNe candidates and
found matches for 9 of our new objects (see Table~\ref{table5} for
details concerning their F$_{12}, {\rm F}_{25}, {\rm F}_{60}, {\rm
F}_{100}$~fluxes and their flux quality density). Taking into account
the fact that the flux quality density for most of them was 1 (1 is
the upper detection limit, whereas 3 is a good quality flux density),
a reasonable number of them satisfy the standard criteria (F$_{12} /
{\rm F}_{25} \leq 0.35$~and F$_{25} / {\rm F}_{60} \geq 0.3$ --
Pottasch et al. 1988; van de Steene 1995) used to accept a new object
as a probable PN. Note that a cross--check for our new PNe was also
performed both with the 2MASS Point Source Catalog (2000), the MSX
infrared astrometric Catalog (Egan et al. 1999) and radio known PNe
(found in the catalogues given in Table 3) in order to find any PNe
counterparts but the results proved negative most likely due to the
low sensitivity limit of these surveys. Thus, radio pointed
observations at selected locations would be needed.

\section{The 1.3 meter telescope Observations}

\subsection{Imaging}

Optical images of the new PNe candidates were also obtained with the
1.3 m (f/7.7) Ritchey-Cretien telescope at Skinakas Observatory during
2001 in May 27--29, June 29--30 and August 10--11 using an \ha
$+$\nitrogen\ interference filter (75\AA~bandwidth). The detector was
a 1024$\times$1024 (with 24$\times$24 $\mu$m $^{2}$ pixels) SITe CCD
with a field of view of 8\arcmin.5 $\times$~8\arcmin.5 and an image
scale of 0\arcsec.5 pixel$^{-1}$. One exposure in the \ha
$+$\nitrogen~of 1800 s and two exposures in the continuum of 180 s
were taken. All new PNe candidates observed with this configuration
can be seen in Fig.~4. Note that they are at the centre of each image
and an arrow points to their exact position. The image size is
150\arcsec on both sides. North is at the top and east to the left. It
should be noticed that taking into account the limited telescope time
available, the follow--up observations were performed in
\ha$+$\nitrogen\ in order to study the morphology of the PNe and also
measure their angular extent.

After finding the PNe candidates, we performed again an astrometric
solution for all images, using the method presented in section
3.2.2. The coordinates for the PNe candidates were now calculated with
better accuracy, due to the 8 times better resolution of this
configuration. Typical rms error in each frame was found to be
$\sim$0\arcsec.3. Hence, the coordinates of all the newly discovered
PNe candidates are given in Table~\ref{table5}.

\subsection{Spectroscopy}

Low dispersion spectra were obtained with the 1.3 m telescope at
Skinakas Observatory during 2001 in June 24--28, July 16--17, 23--25
and August 12, 20--25. A 1300 line mm$^{-1}$~grating was used in
conjunction with a 2000$\times$800 SITe CCD (15$\times$15 $\mu$m$^{2}$
pixels) which resulted in a dispersion of 1 \AA\ pixel$^{-1}$~and
covers the range of 4750\AA\ -- 6815\AA. The slit width was 7\arcsec.7
and it was oriented in the south--north direction, while the slit
length was 7\arcmin.9. The exposure time of the individual spectra
ranges from 2400 to 3600 s depending on the observing window of each
object. Two spectra of the same object were obtained whenever possible
and in this case the S/N weighted average line flux values are given.
The spectrophotometric standard stars HR5501, HR7596, HR9087, HR718,
and HR7950 (Hamuy et al. 1992; 1994) were observed in order to
calibrate the spectra of the new PNe candidates.  The low resolution
spectra were taken on the relatively bright optical part of each PN
candidate. Some typical spectra of our new PNe can be seen in
Fig.~\ref{spectra}. In Table~\ref{table7}, we present their line
fluxes corrected for atmospheric extinction and interstellar
reddening, using the scale of F(\hbeta)$=$100. Interstellar reddening
was derived from the
\ha/\hbeta\ ratio (Osterbrock 1989), using the interstellar extinction
law by Fitzpatrick (1999) and R$_{\rm v} = 3.1$~for all
objects. Therefore, the interstellar extinction c(\hbeta) can be
derived by using the relationship

\begin{equation}
{\rm c(H\beta)} = \frac{1}{0.348} \log\frac{{\rm F(H\alpha)}/{\rm
F(H\beta)}}{2.85}
\end{equation} 

where, 0.348 is the relative logarithmic extinction coefficient for
\hbeta/\ha\ and 2.85 the theoretical value of F(\ha)/F(\hbeta).
The observational reddening in magnitude E$_{\rm B-V}$~was also
calculated using the relationship (Seaton 1979)

\begin{equation}
{\rm c(H\beta) = 0.4 \cdot X_{\beta} \cdot E_{\rm B-V}}
\end{equation} 

where the extinction parameter X$_{\beta} =$ 3.615 (Fitzpatrick
1999). The 1$\sigma$ error of the extinction c(\hbeta) and E$_{\rm
B-V}$~was calculated through standard error propagation of equations
(1) and (2).  Because of the high interstellar extinction in the
direction of the bulge a second estimation of E$_{\rm B-V}$ was made
using the SFD code (Schlegel, Finkbeiner \& Davis 1998) which uses
infrared maps to estimate Galactic extinction. A comparison shows a
general agreement between the two E$_{\rm B-V}$~calculations within a
3$\sigma$ error. The signal to noise ratios do not include calibration
errors, which are less than 10 percent. The absolute \ha\ fluxes (in
units of erg s$^{-1}$ cm$^{-2}$ arcsec$^{-2}$), the exposure time of
each individual spectra, the interstellar extinction c(\hbeta) with
its estimated error and the reddening E$_{\rm B-V}$~(resulting from
our observations and SFD maps) are listed in Table~\ref{table8}. Note
that due to the low S/N ratio ($<$ 10) of the
\hbeta\ line for some of the new PNe candidates, their accuracy is
lower while, the spectrum of PTB6 was taken during a non photometric
night, thus its accuracy is even lower and only its relative line
fluxes should be taken into account.

\section{Discussion}

The \ha $+$\nitrogen\ images as well as the low resolution spectra of
the newly discovered PNe candidates were used for a more detailed
study.  Their relative emission line fluxes confirm that all are
photoionized nebulae, according to the diagnostic diagram
log(\ha/\nitrogen) vs. log(\ha/\sulfurt) (Garc\'{\i}a et al. 1991),
where they spread well inside the area occupied by PNe (Fig. 3) and
their \sulfurt/\ha\ ratio which is in all cases less than 0.3. Note that
five of them have very low S/N ratio in their \nitrogen\ and \sulfurt\
emission lines which are not presented in Table 7. However, their
\sulfurt/\ha\ is less than 0.2. The angular diameters of the newly 
discovered PNe have been measured according their optical images
(\ha$+$\nitrogen), taken with the 1.3 m telescope. It should be
noticed that (a) the diameters were defined from the optical outer
part of the nebulae, but in some cases it might be slightly larger if
a very faint halo exists and (b) that in the case of elliptical shell
we present both the major and the minor axes. All diameters are given
in Table~\ref{table8}. Thirteen of the new PNe have diameter
$\leq$20\arcsec\ (Bulge limit -- Gathier et al. 1983), 4 are close
to this limit ($\leq$25\arcsec), while 8 have diameter
$\geq$30\arcsec. The possibility that the latter are evolved
and/or their local nature cannot be excluded.  

In most cases, a spherical well-defined shell (star-like or ring) was
found but nonspherical nebulae (e.g. elliptical) are also present. In
particular, 50 percent of the new PNe have a ring--like structure
(e.g. PTB15, PTB17), while the rest are divided between faint nebulae
with incomplete bright shells (e.g. PTB3, PTB23) and bright compact
nebulae (e.g. PTB5, PTB11).  The morphological differences could also
be attributed to the PN stage of evolution. Compact nebulae in their
first stage of expansion look stellar and relatively bright, while
large nebulae with faint surface brightness are in their late
evolutionary stage. However, three of the new PNe look different. PTB4
is an elongated ring--like nebula probably due to its position angle
while, PTB19 is also elongated but with non spherical symmetry at all
and PTB7 seems to have a unusual bipolar shape, thus it would be
interesting to investigate its morphology further in the future. A
morphological type was assigned to the new PNe (Table~\ref{table8})
according to Manchado et al. (2000).

\begin{table*}
\centering
\caption[]{Line fluxes.}
\label{table7}
\begin{tabular}{llllllll}
\hline
Line (\AA) & PTB1 & PTB2 & PTB3 & PTB4 & PTB5$^{\rm a}$ & PTB6 & PTB7 \\
\hline
\hbeta\ 4861 & 100.0 (5)$^{\rm b}$ & 100.0 (11) & 100.0 (6) & 100.0 (8) & 100.0 (39) & 100.0 (3) & 100.0 (11) \\
\oxygen\ 4959 & 317.5 (13) & 313.4 (33) & 183.8 (12) & 213.2 (18) & 425.0 (111) & 43.1 (2) & 172.6 (21)  \\
\oxygen\ 5007 & 851.3 (32) & 928.3 (77) & 507.3 (32) & 587.0 (47) & 1254.2 (184) & 405.6 (5) & 513.9 (58) \\ 
\heliumb\ 5411 & $-$ & $-$ & $-$ & 4.1 (2) & 4.7 (9) & $-$ & 12.8 (6) \\
\nitrogen\ 5755 & $-$ & $-$ & $-$ & $-$ & $-$ & $-$ & $-$ \\
\helium\ 5876 & 17.8 (7) & 9.1 (7) & 23.2 (8) & 16.3 (11) & 10.3 (32) & 47.1 (4) &  17.5 (10) \\
\heliumb\ 6234 & $-$ & $-$ & $-$ & $-$ & 0.3 (8) & $-$ & $-$\\
\nitrogen\ 6548 & $-$ & $-$ & 33.3 (21) & 41.0 (29) & 1.5 (10) & 91.3 (9) & 2.7 (4)  \\
\ha\ 6563 & 285.0 (78) & 285.0 (141) & 285.0 (104) & 285.0 (113) & 285.0 (309) & 285.0 (24) & 285.0 (139) \\
\nitrogen\ 6584 & $-$ & $-$ & 104.1 (57) & 126.0 (74)  & 4.5 (24) & 205.8 (23) & 12.1 (9)  \\
\helium\ 6678 & 3.6 (8) & 1.7 (6) & 8.4 (22) & 5.2 (7) & 3.0 (21) & $-$ & 2.6 (2) \\
\sulfurt\ 6716 & 5.7 (12) & $-$ & 9.8 (11) & 16.0 (17) & 0.9 (7) & 21.9 (2) & 3.9 (5) \\
\sulfurt\ 6731 & 6.3 (17) & $-$ & 7.4 (9) & 12.2 (15) & 1.2 (12) & 16.8 (2) & 2.2 (4) \\
\hline
Line (\AA) &  PTB8 & PTB9 & PTB10 & PTB11 & PTB12 & PTB13 & PTB14  \\
\hline
\hbeta\ 4861 &  100.0 (5) & 100.0 (23) & 100.0 (8) & 100.0 (12) & 100.0 (21) & 100.0 (3) & 100.0 (24) \\
\oxygen\ 4959 & 327.1 (15) & 117.2 (39) & 316.6 (27) & 305.0 (40) & 214.5 (48) & 384.9 (7) & 345.5 (70)  \\
\oxygen\ 5007 & 987.5 (41) & 357.5 (90) & 954.3 (63) & 923.4 (87) & 629.4 (105) & 1573.1 (21) & 1040.5 (142)  \\ 
\heliumb\ 5411 & $-$ & 8.6 (12) & $-$ & $-$ & 5.5 (5) & 20.0 (2) & $-$ \\
\nitrogen\ 5755 & $-$ & $-$ & $-$ & 4.6 (2) & $-$ & $-$ & $-$ \\
\helium\ 5876 & 11.0 (4) & 29.1 (33) & $-$ & 19.3 (16) & 10.4 (19) & 30.1 (8) & 18.3 (31) \\
\heliumb\ 6234 & $-$ & $-$ & $-$ & $-$ & $-$ & 9.2 (3) & $-$ \\
\nitrogen\ 6548 & 16.7 (9) & 7.7 (16) & $-$ & 95.4 (89) & 7.4 (11) & $-$ & 1.0 (3) \\
\ha\ 6563 & 285.0 (71) & 285.0 (240) & 285.0 (130) & 285.0 (175) & 285.0 (129) & 285.0 (52) & 285.0 (258)  \\
\nitrogen\ 6584 &  51.5 (22) & 29.9 (48) & $-$ & 293.9 (174) & 24.7 (26) & $-$ & 5.0 (12)  \\
\helium\ 6678 & 2.0 (2) & 4.9 (25) & $-$ & 4.7 (12) & 3.3 (7) & $-$ & 4.2 (17) \\
\sulfurt\ 6716 & 7.0 (5) & 3.7 (17) & $-$ & 65.7 (82) & 6.7 (15) & $-$ & 0.5 (3) \\
\sulfurt\ 6731 & 7.9 (6) & 4.3 (19) & $-$ & 58.7 (77) & 5.9 (14) & $-$ & 0.5 (1) \\
\hline
Line (\AA) & PTB15 & PTB16 & PTB17$^{\rm a}$ & PTB18$^{\rm a}$ & PTB19 & PTB20 & PTB21  \\
\hline
\hbeta\ 4861 &  100.0 (6) & 100.0 (6) & 100.0 (13) & 100.0 (5) & 100.0 (26) & 100.0 (12) & 100.0 (2) \\
\oxygen\ 4959 & 235.3 (14) & 136.1 (10) & 203.4 (29) & 204.3 (13) & 222.0 (56) & 471.0 (47) & 431.1 (7)\\
\oxygen\ 5007 & 653.1 (38) & 462.8 (29) & 604.5 (71) & 600.4 (36) & 646.7 (110) & 1412.3 (90) & 949.6 (18) \\ 
\heliumb\ 5411 & 11.6 (3) & 10.2 (3) & $-$ & 5.6 (9) & $-$ & 4.9 (5) & $-$ \\
\nitrogen\ 5755 & $-$ & $-$ & 4.0 (4) & $-$ & $-$ & $-$ & $-$ \\
\helium\ 5876 & $-$ & 16.6 (7) & 22.4 (18) & 12.9 (8) & 16.3 (41) & 16.6 (16) &$-$ \\
\heliumb\ 6234 & 2.4 (5) & $-$ & $-$ & $-$ & $-$ & $-$ & $-$ \\
\nitrogen\ 6548 & $-$ & 2.6 (2) & 103.6 (57) & 88.4 (19) & 2.6 (12) & 9.2 (26) & 63.0 (6)  \\
\ha\ 6563 & 285.0 (82) & 285.0 (83) & 285.0 (94) & 285.0 (44) & 285.0 (198) & 285.0 (192) & 285.0 (22) \\
\nitrogen\ 6584 & $-$ & 7.8 (6) & 323.3 (77) & 270.5 (36) & 8.2 (30) & 32.7 (22) & 172.7 (14) \\
\helium\ 6678 & $-$ & 10.9 (3) & 5.0 (9) & 6.6 (2) & 4.4 (28) & 3.3 (14) & 10.9 (3)  \\
\sulfurt\ 6716 & $-$ & 7.4 (2) & 36.9 (28) & 41.1 (16) & 1.1 (9) & 1.2 (9) & 40.1 (5)  \\
\sulfurt\ 6731 & $-$ & 4.5 (1) & 29.1 (23) & 33.2 (12) & 1.4 (11) & 2.5 (11) & 27.1 (4) \\
\hline
Line (\AA) & PTB22 & PTB23$^{\rm a}$ & PTB24 & PTB25$^{\rm a}$ \\
\hline
\hbeta\ 4861 &  100.0 (4) & 100.0 (28) & 100.0 (7) & 100.0 (19) \\
\oxygen\ 4959 & 380.7 (13) & 397.4 (88) & 260.7 (19) & 231.4 (47) \\
\oxygen\ 5007 & 1147.6 (38) & 1152.2 (172) & 838.4 (49) & 671.6 (111) \\ 
\heliumb\ 5411 & $-$ & 9.6 (8) & $-$ & 8.1 (8) \\
\nitrogen\ 5755 & $-$ & $-$ & 3.3 (9) & $-$ \\
\helium\ 5876 & $-$ & 2.3 (4) & 14.0 (8) & 3.0 (5) \\
\heliumb\ 6234 & $-$ & $-$ & $-$ & 1.2 (5) \\
\nitrogen\ 6548 & $-$ & 2.0 (7) & 76.9 (36) & 1.9 (5) \\
\ha\ 6563 & 285.0 (45) & 285.0 (147) & 285.0 (97) & 285.0 (207) \\
\nitrogen\ 6584 & $-$ & 5.7 (14) & 233.9 (82) & 7.0 (14) \\
\helium\ 6678 & $-$ & 1.5 (3) & 3.2 (3) & $-$ \\
\sulfurt\ 6716 & $-$ & 3.1 (10) & 40.4 (27) & 3.3 (8) \\
\sulfurt\ 6731 & $-$ & 3.5 (8) & 29.2 (20) & 3.1 (7) \\
\hline
\end{tabular}

\medskip{}
\begin{flushleft}

${\rm ^a}$ Listed fluxes are a signal to noise weighted average of the individual fluxes\\ 
${\rm ^b}$ Numbers in parentheses represent the signal to noise ratio of the line fluxes, measured at the center of the corresponding emission line profile.\\ 

${\rm }$ All fluxes normalized to F(\hbeta)=100 and they are corrected for interstellar extinction.\\
\end{flushleft}
\end{table*}

\begin{table*}
\centering
\caption{Exposure time and basic physical parameters.}
\label{table8}
\begin{tabular}{c|c|c|c|c|c|c|c|c|c|c|c}
\hline
Object & Exp. time & Diameter$^{\rm 1}$ & M.Type$^{\rm 2}$ & F(\ha)$^{\rm 3}$ & c(\hbeta)$^{\rm 4}$ & $\sigma_{\rm c(H\beta)}^{\rm 5}$ & E$_{\rm B-V (OBS)}^{\rm 6}$ & $\sigma_{\rm E_{B-V}}^{\rm 7}$  & E$_{\rm B-V (SFD)}^{\rm 8}$  &n$_{\rm \sulfurt}^{\rm 9}$ &  $\sigma_{\rm n_{\rm [S~{\sc ii}]}}^{\rm 10}$\\
\hline
PTB1 & 3600$^{\rm 11}$ (1)$^{\rm 12}$ & 33.0 & R & 3.9 & 2.25 & 0.26 & 1.55 & 0.18 & 1.26 & 0.93 & 0.33 \\
PTB2 & 3600 (1) & 15.0 & B & 11.1 & 1.38 & 0.12 & 0.95 & 0.08 & 1.18 & $<$0.07 & $-$ \\
PTB3 & 3600 (1) & 23.0 & R & 4.52 & 1.78 & 0.21 & 1.23 & 0.15 & 0.92 & 0.09 & 0.08 \\
PTB4 & 2700 (1) & 17.0$\times$8.0 & E & 9.0 & 1.70 & 0.15 & 1.17 & 0.10 & 1.09 & 0.12 & 0.10 \\
PTB5 & 4800 (2) & 10.0 & R & 40.7 & 1.20 & 0.02 & 0.83 & 0.01 & 0.84 & 1.88 & 0.81 \\
PTB6 & 2700 (1) & 20.0 & E & 1.2 & 1.42 & 0.50 & 0.98 & 0.35 & 0.93 & $<$0.07 & $-$ \\
PTB7 & 3600 (1) & 25.0$\times$10.0 & B & 4.6 & 0.85 & 0.11 & 0.59 & 0.08 & 0.64 & $<$0.07 & $-$ \\
PTB8 & 2500 (1) & 19.0 & R & 5.6 & 1.73 & 0.25 & 1.19 & 0.17 & 1.10 & 0.95 & 0.88 \\
PTB9 & 3600 (1) & 18.0$\times$16.0 & E & 5.9 &  1.25 & 0.05 & 0.86 & 0.04 & 0.92 & 1.10 & 0.31 \\
PTB10 & 3600 (1) & 10.0$\times$8.0 & E & 7.2 & 1.84 & 0.16 & 1.27 & 0.11 & 0.94 & $<$0.07 & $-$ \\
PTB11 & 3600 (1) & 10.5 & R & 17.6 & 1.92 & 0.11 & 1.32 & 0.07 & 1.30 & 0.36 & 0.03 \\
PTB12 & 3600 (1) & 17.0$\times$14.0 & E & 15.7 & 1.19 & 0.06 & 0.82 & 0.04 & 0.90 & 0.34 & 0.18 \\
PTB13 & 3600 (1) & 11.0 & R & 4.3 & 2.74 & 0.48 & 1.89 & 0.33 & 1.05 & $<$0.07 & $-$ \\
PTB14 & 3600 (1) & 9.0 & R & 32.5 &  1.58 & 0.05 & 1.09 & 0.03 & 0.92 & 0.28 & 0.27 \\
PTB15 & 3600 (1) & 33.0 & R & 4.2 & 1.58 & 0.21 & 1.09 & 0.15 & 1.12 & $<$0.07 & $-$ \\
PTB16 & 3600 (1) & 24.5 & R & 3.3 & 1.08 & 0.22 & 0.74 & 0.15 & 0.51 & $<$0.07 & $-$ \\
PTB17 & 5400 (2) & 32.0 & R & 9.0 & 1.35 & 0.07 & 0.93 & 0.05 & 0.92 & 0.16 & 0.06 \\
PTB18 & 4800 (2) & 38.0 & R & 2.2 & 0.63 & 0.19 & 0.44 & 0.13 & 0.51 & 0.19 & 0.11 \\
PTB19 & 3600 (1) & 17$\times$20.0 & E & 40.1 & 1.93 & 0.05 & 1.33 & 0.03 & 1.31 & 1.71 & 0.72 \\
PTB20 & 2700 (1) & 10.0 & R & 7.8 & 1.94 & 0.11 & 1.34 & 0.08 & 1.25 & 20.74 & 12.0 \\
PTB21 & 3600 (1) & 69.0 & R & 0.6 & 0.94 & 0.48 & 0.65 & 0.33 & 1.07 & $<$0.07 & $-$ \\
PTB22 & 2700 (1) & 34.5 & R & 2.1 & 1.48 & 0.34 & 1.02 & 0.23 & 0.98 & $<$0.07 & $-$ \\
PTB23 & 4800 (2) & 46.0$\times$38.0 & E & 24.8 & 0.62 & 0.13 & 0.43 & 0.09 & 1.69 & 2.05 & 0.60\\
PTB24 & 2700 (1) & 16.0 & R & 4.3 & 1.55 & 0.19 & 1.07 & 0.13 & 1.13 & $<$0.07 & $-$ \\
PTB25 & 5400 (2) & 40.0 & R & 9.2 & 0.99 & 0.15 & 0.69 & 0.10 & 1.44 & 0.47 & 0.32 \\
\hline
\end{tabular}

\medskip{}
\begin{flushleft}

$^{\rm 1}$ Optical diameters in arcsec. \\ 

$^{\rm 2}$ Morphological classification according to Manchado et al. (2000), where R($=$round), E ($=$elliptical) and B ($=$bipolar).\\ 

$^{\rm 3}$ Absolute \ha\ flux in units of \flux. \\

$^{\rm 4}$ Logarithmic extinction at \hbeta\ (see Sect. 4.2). \\

$^{\rm 5}$ 1$\sigma$~error on the logarithmic extinction. \\ 

$^{\rm 6}$ Observed Reddening (see Sect. 4.2). \\

$^{\rm 7}$ Error on the observed reddening.\\

$^{\rm 8}$ Reddening according the maps of Schlegel et al. 1998 (see Sect. 4.2). \\

$^{\rm 9}$ Electron density in units of 10$^{3}$~cm$^{-3}$ (see Sect. 5).\\ 

$^{\rm 10}$ Error on the electron density.\\

$^{\rm 11}$ Total exposure time in sec. \\

$^{\rm 12}$ Number of individual exposures. \\

\end{flushleft}
\end{table*}

Assuming a temperature of 10$^{4}~^{o}$K, it is possible to estimate
the electron density n$_{\rm \sulfurt}$, according to the task
``temden'' in the nebular package in IRAF for a specific sulfur line
ratio (\sulfurt\ 6716\AA/\sulfurt\ 6731\AA). This task computes the electron density
given an electron temperature, of an ionized nebular gas within the
5-level atom approximation (Shaw \& Dufour 1995). Furthermore the
error of the electron density was calculated through standard error
propagation. The electron densities and their associated errors are
given in Table~\ref{table8}. Using the observed \nitrogen\
6548$+$6584\AA/\nitrogen\ 5755\AA, an estimation for the electron temperature T$_{\rm
\nitrogen}$~was made whenever possible. In fact only for three of our
new PNe (PTB11, PTB17 \& PTB24) values close to 10000 $^{o}$K were
found. In those cases where a temperature measurement was not
available we assumed a temperature of 10$^{4}~^{o}$K (Cuisinier et
al. 2000). The high interstellar extinction towards the Galactic bulge
as well as the low sensitivity brightness of the new PNe are the main
reasons which do not allow to record high quality spectra. This
affects mainly the lower intensity lines like \sulfurt\ or \heliumb,
which are faint or even undetectable. For all PNe of our survey, \ha\
and \oxygen\ emission lines could be measured in good conditions and
in most cases, \hbeta\ and \NII. However, the spectrum for some of the
new PNe (e.g. PTB22) has a low signal to noise ratio and only \hbeta,
\ha\ and \oxygen\ are well determined. As can be seen from Fig. 7
emission from \oiii\ is present in all observed spectra. Since these
PNe are found in a region of significant extinction, shorter
wavelengths are more affected than longer wavelengths. In Table 5 we
list the measured fluxes in our long--slit spectra corrected for
interstellar extinction and is evident that almost all PNe display
very strong \oiii\ emission relative to \ha. It is known that the N/O
and He abundances reflect the mass distribution of the progenitor
star, with more massive objects having higher N/O and He abundances in
comparison with the small--mass objects (Cuisinier et al. 2000). In
our case, the abundances in N/O and He generally showed low N/O and
He, implying old progenitor stars. The latter suggests that they
belong to the Galactic bulge region according to Webster (1988) and
Cuisinier et al. (2000) who both showed that the Galactic bulge PNe
have old progenitors with low N/O and He abundances. Most of our PNe
seem to be of low excitation since they display \helium\ lines but not
\heliumb. However, there are a few that emit in \heliumb\ but not in
\helium\ (like PTB15) which must be highly ionized and this may
explain the weakness or absence of the \NII\ lines. The excitation
classes of our new PNe were studied according the classification
criteria of Aller (1956), Feast (1968) and Webster (1975). They are
derived basically from the ratios of \oxygen/\hbeta, \heliumb\
4686 \AA/\hbeta, \oii\ 3727 \AA/\oxygen\ 4959 \AA\ and
\heliumb\ 4686 \AA/\helium\ 5876 \AA. In our case, the
\oxygen/\hbeta\ ratio suggests that our new PNe belong to the medium
excitation classes (4--6). However, the absence of the \heliumb\ 4686 \AA\
and \oii\ 3727 \AA\ emission lines from our spectra do not allow us to
better confine the actual excitation class of each PN.

\section{Conclusion}

We present the first results of a narrowband \oiii\ survey for PNe in
the Galactic bulge ($l>0^{o}$). Covering 66 percent of our selected
region, we detected 90 objects, including 57 known PNe, 8 known PNe
candidates and 25 new PNe. \ha $+$\nitrogen\ images as well as low
resolution spectra of the new PNe were taken and confirmed the
photoionized nature of the emission. The 70 percent of them have
angular sizes $\leq$ 20\arcsec--25\arcsec\ while all show \ha/\oiii
$<$ 1. Nine (9) of our new PNe are found to be associated with IRAS
sources. They all display low N/O and He abundances implying old
progenitor stars, which is one of the characteristics of the Galactic
bulge PNe.

\section*{Acknowledgments} 

The authors would like to thank the referee for his comments and
suggestions which helped to clarify, and enhance the scope of this
paper. We would also like to thank Q. Parker and A. Acker who provided
us the Edinburgh/AAO/Strasbourg Catalogue of Galactic Planetary
Nebulae and the First Supplement to the Strasbourg--ESO Catalogue of
Galactic Planetary Nebulae, respectively and E. Semkov, G. Paterakis,
A. Kougentakis and A. Strigachev for their assistance and help during
these observations. Skinakas Observatory is a collaborative project of
the University of Crete, the Foundation for Research and
Technology-Hellas, and the Max-Planck-Institut f\"{u}r
extraterrestrische Physik.

\bibliographystyle{mnras}

\begin{thebibliography}{}

\bibitem[\protect\citename{Acker et al}1992a]{bu}
Acker A., Cuisinier F., Stenholm B. \& Terzan A., 1992a, A\&A, 264, 217

\bibitem[\protect\citename{Acker et al}1992b]{bu}
Acker A., Ochsenbein F., Stenholm B., Tylenda R., Marcout J. \& Schohn
C., 1992b, in the Strasbourg-ESO Catalogue of Galactic Planetary
Nebulae, Parts 1 and 2 (Strasbourg: ESO)

\bibitem[\protect\citename{Acker et al}1996]{bu}
Acker A., Marcout J., Ochsenbein F., Beaulieu S., Garcia-Lario P. \&
Jacoby G., 1996, First Supplement to the Strasbourg-ESO Catalogue of
Galactic Planetary Nebulae (Strasbourg: Obs. Strasbourg)

\bibitem[\protect\citename{Aller}1956]{bu}
Aller L. H., 1956, Gaseous Nebulae (New York: Wiley)

\bibitem[\protect\citename{Beaulieu et al}1999]{bu}
Beaulieu S. F., Dopita M. A. \& Freeman K. C., 1999, ApJ, 515, 610

\bibitem[\protect\citename{Beaulieu et al}2000]{bu}
Beaulieu S. F., Freeman K. C., Kalnajs A. J., Saha P. \& Zhao H.,
2000, AJ, 120, 855

\bibitem[\protect\citename{Boumis \& Papamastorakis}2002]{bu}
Boumis P. \& Papamastorakis J., 2002, proceedings of the 5th Hellenic
Astronomical Society Conference, held in Fodele Crete, Greece, 20-22
September 2001, in press

\bibitem[\protect\citename{Boumis et al}2002]{bu}
Boumis P. et al., 2003, MNRAS, in preparation (paper II)

\bibitem[\protect\citename{Cappellaro et al}2001]{bu}
Cappellaro E., Sabbadin F., Benetti S. \& Turatto M., 2001, A\&A, 377, 1035

\bibitem[\protect\citename{Ciardullo et al}1989]{bu}
Ciardullo R., Jacoby G. H., Ford H. C. \& Neil J. D., 1989, ApJ, 339, 53

\bibitem[\protect\citename{Condon et al}1999]{bu}
Condon J. J., Kaplan D. L. \& Terzian Y., 1999, ApJS, 123, 219

\bibitem[\protect\citename{Cuisinier et al}2000]{bu}
Cuisinier F., Maciel W. J., K\"{o}ppen J., Acker A. \& Stenholm B.,
2000, A\&A, 353, 543

\bibitem[\protect\citename{Durand et al}1998]{bu}
Durand S., Acker A. \& Zijlstra A., 1998, A\$AS, 132, 13

\bibitem[\protect\citename{Egan et al}1999]{bu}
Egan, M. P., et al., 1999, in Astrophysics with Infrared Surveys, A
Prelude to SIRTF, ed. M. D. Bicay, R. M. Cutri, \& B. F. Madore, ASP
Conf. Ser., 177, 404

\bibitem[\protect\citename{Escudero \& Costa}2001]{bu}
Escudero A. V. \& Costa R. D. D., 2001, A\&A, 380, 300

\bibitem[\protect\citename{Feast}1987]{bu}
Feast M. W., 1987, The Galaxy, eds. Gilmore G. \& Carswell B. (Reidel,
Dordrecht), 1

\bibitem[\protect\citename{Feldmeier et al}1997]{bu}
Feldmeier J. J., Ciardullo R. \& Jacoby G. H., 1997, ApJ, 479, 231

\bibitem[\protect\citename{Fitzpatrick}1999]{bu}
Fitzpatrick E. L., 1999, PASP, 111, 63

\bibitem[\protect\citename{Garcia et al}1991]{bu} Garc\'{\i}a-Lario P.,
Manchado A., Riera A., Mampaso A. \& Pottasch S. R., 1991, A\&A, 249,
223

\bibitem[\protect\citename{Gathier et al}1983]{bu}
Gathier R., Pottasch S. R., Goss W. W. \& van Gorkom J. H., 1983,
A\&A, 128, 325

\bibitem[\protect\citename{Hamuy et al}1992]{bu} 
Hamuy M., Walker A. R., Suntzeff N. B., Gigoux P., Heathcote S. R. \&
Phillips M. M., 1992, PASP, 104, 533

\bibitem[\protect\citename{Hamuy et al}1994]{bu}
Hamuy M., Suntzeff N. B., Heathcote S. R., Walker A. R., Gigoux P. \&
Phillips M. M., 1994, PASP, 106, 566

\bibitem[\protect\citename{Hui et al}1993]{bu}
Hui X., Ford H. C., Ciardullo R. \& Jacoby G. H., 1993, ApJ, 414, 463

\bibitem[\protect\citename{IRAS}1988]{bu}
IRAS Point Source Catalog, Version 2, 1988, Joint IRAS Science Working
Group (Washington, DC: GPO)

\bibitem[\protect\citename{Jacoby et al}1990]{bu}
Jacoby G. H., Ciardullo R., Ford H. C. \& Booth J., 1990, ApJ, 365, 471

\bibitem[\protect\citename{Kimeswenger et al}1997]{bu}
Kimeswenger S., Kienel C. \& Widauer H., 1997, Ap\&SS, 210, 105

\bibitem[\protect\citename{Kimeswenger}2001]{bu}
Kimeswenger S., 2001, Rev.Mex.A.A, 37, 115

\bibitem[\protect\citename{Kinman et al}1988]{bu}
Kinman T. D., Feast M. W. \& Lasker B. M., 1988, AJ, 95, 804

\bibitem[\protect\citename{Kohoutek}1997]{bu}
Kohoutek L., 1997, AN, 318, 35

\bibitem[\protect\citename{Kohoutek}2001]{bu}
Kohoutek L., 2001, A\&A, 378, 843

\bibitem[\protect\citename{K\"{o}ppen \& Vergely}1999]{bu}
K\"{o}ppen J. \& Vergely J.-L., 1998, MNRAS, 299, 567

\bibitem[\protect\citename{Lasker et al}1999]{bu}
Lasker B. M., Russel J. N. \& Jenkner H., 1999, in the HST Guide Star
Catalog, version 1.1-ACT, The Association of Universities for Research
in Astronomy, Inc

\bibitem[\protect\citename{Lauberts}1982]{bu}
Lauberts A., 1982, The ESO/Uppsala Survey of the ESO (B) Atlas, European Southern Observatory

\bibitem[\protect\citename{2MASS}2000]{bu}
2MASS Point Source Catalog, 2000, NASA/IPAC Infrared Science Archive,
Jet Propulsion Laboratory (California Institute of Technology)

\bibitem[\protect\citename{Manchado et al}1996]{bu}
Manchado A., Guerrero M. A., Stanghellini L. \& Serra-Ricart M., 1996,
in the IAC Morphological Catalog of Northern Galactic Planetary
Nebulae (IAC)

\bibitem[\protect\citename{Manchado et al}2000]{bu}
Manchado A., Villaver E., Stanghellini L. \& Guerrero M., 2000, in ASP
Conf. Ser. 199, Asymmetrical Planetary Nebulae II. From Origins to
Microstructures, eds. Kastner et al., 17

\bibitem[\protect\citename{Minniti et al}1995]{bu}
Minniti D., Olszewski E. W., Liebert J., White S. D. M., Hill J. M. \&
Irwin M. J., 1995, MNRAS, 277, 1293

\bibitem[\protect\citename{Minniti}1996a]{bu}
Minniti D., 1996a, ApJ, 459, 175

\bibitem[\protect\citename{Minniti}1996b]{bu}
Minniti D., 1996b, ApJ, 459, 579

\bibitem[\protect\citename{Moreno et al}1988]{bu}
Moreno H., Lasker B. M., Gutierrez-Moreno A. \& Torres C., 1988, PASP, 100, 604

\bibitem[\protect\citename{Morison \& Harding}1993]{bu}
Morison H. L. \& Harding P., 1993, PASP, 105, 977

\bibitem[\protect\citename{Osterbrock}1989]{bu}
Osterbrock D. E., 1989, Astrophysics of Gaseous Nebulae and Active
Galactic Nuclei, vol. 9 (University Science Books)

\bibitem[\protect\citename{Parker et al}2001a]{bu}
Parker Q. A., Hartley M., Russeil D., Acker A., Morgan D., Beaulieu
S., Morris R., Phillips S. \& Cohen M., 2001a, IAU Symposium No. 209,
Planetary Nebulae: Their Evolution and Role in the Universe (Canberra,
Australia, November 19 - 23 2001), in press

\bibitem[\protect\citename{Parker et al}2001b]{bu}
Parker Q. A., Hartley M., Russeil D., Acker A., Ochsenbein F., Morgan
D., Beaulieu S., Morris R., Marcout J., Cohen M. \& Phillips S.,
2001b, in the Edinburgh/AAO/Strasbourg Catalogue of Galactic Planetary
Nebulae, Preliminary version 1.0 November 2001

\bibitem[\protect\citename{Pottasch et al}1988]{bu}
Pottasch S. R., Bignell C., Olling R. \& Zijlstra A. A., 1988, A\&A, 205, 248

\bibitem[\protect\citename{Ratag}1990]{bu}
Ratag M. A., 1990, PhD thesis, Univ. Groningen

\bibitem[\protect\citename{Sevenster}1999]{bu}
Sevenster M. N., 1999, MNRAS, 310, 629

\bibitem[\protect\citename{Schlegel et al}1998]{bu}
Schlegel D. J., Finkbeiner D. P. \& Davis M., 1998, ApJ, 500, 525

\bibitem[\protect\citename{Schneider \& Buckley}1996]{bu}
Schneider S. E. \& Buckley D., 1996, ApJ, 459, 606

\bibitem[\protect\citename{Seaton}1979]{bu}
Seaton M. J., 1979, MNRAS, 187, L73

\bibitem[\protect\citename{Shaw \& Dufour}1995]{bu}
Shaw R. A. \& Dufour R. J., 1995, PASP, 107,896

\bibitem[\protect\citename{van de Steene}1995]{bu}
van de Steene G. C., 1995, PhD thesis, Univ. Groningen

\bibitem[\protect\citename{van de Steene \& Jacoby}2001]{bu}
van de Steene G. C., \& Jacoby G. H., 2001, A\&A, 373, 536

\bibitem[\protect\citename{van Hoof \& van de Steene}1999]{bu}
van Hoof P. A. M. \& van de Steene G. C., 1999, 308, 623

\bibitem[\protect\citename{Walker \& Terndrup}1991]{bu}
Walker A. R. \& Terndrup D. M., 1991, ApJ, 378, 119

\bibitem[\protect\citename{Webster}1975]{bu}
Webster L., 1975, MNRAS, 173, 437

\bibitem[\protect\citename{Webster}1988]{bu}
Webster L., 1988, MNRAS, 230, 377

\bibitem[\protect\citename{Weiland et al}1993]{bu}
Weiland J. L., et al., 1993, in AIP Conf. Proc. 278, Back to the
Galaxy, eds. Holt S.S. \& Verter F. (New York: AIP), 137

\bibitem[\protect\citename{Whitelock}1993]{bu}
Whitelock P., 1993, in IAU Symp. 153, Galactic Bulges, eds. Dejonghe
H. \& Habing H. J. (Dordrecht: Kluwer), 39

\bibitem[\protect\citename{Whitelock et al}1994]{bu}
Whitelock P., Menzies J., Feast M., Marang F., Carter B., Roberts G.,
Catchpole R. \& Chapman J., 1994, MNRAS, 267, 711

\bibitem[\protect\citename{Xilouris et al}1994]{bu}
Xilouris K. M., Papamastorakis J., Sokolov N., Paleologou E. \& Reich
W., 1994, A\&A, 290, 639

\bibitem[\protect\citename{Zijlstra \& Pottasch}1991]{bu}
Zijlstra A. A. \& Pottasch S. R., 1991, A\&A, 243, 478

\end{thebibliography}

\begin{figure*}
\centering
\mbox{\epsfxsize=6in\epsfbox[30 200 554 687]{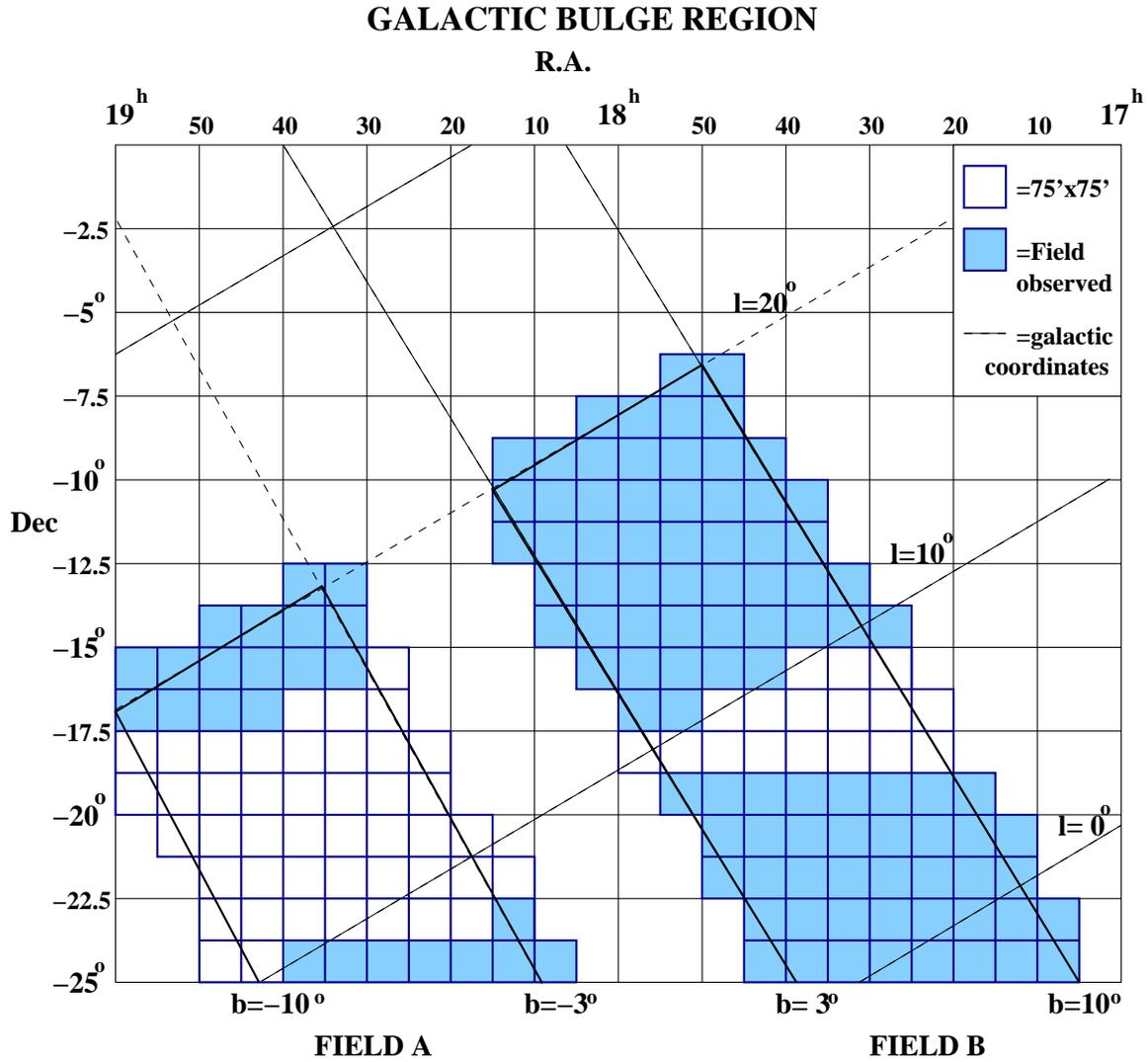}}
\caption[]{Optical imaging survey grid in equatorial
coordinates. Galactic coordinates are also included (dash lines) to
permit an accurate drawing of the selected Bulge region (bold solid
lines). The filled and open rectangles represent the observed and not
yet observed fields in the year 2000, respectively.}
\label{fig01}
\end{figure*}

\begin{figure*}
\centering
\mbox{\epsfxsize=6.6in\epsfbox[20 20 575 762]{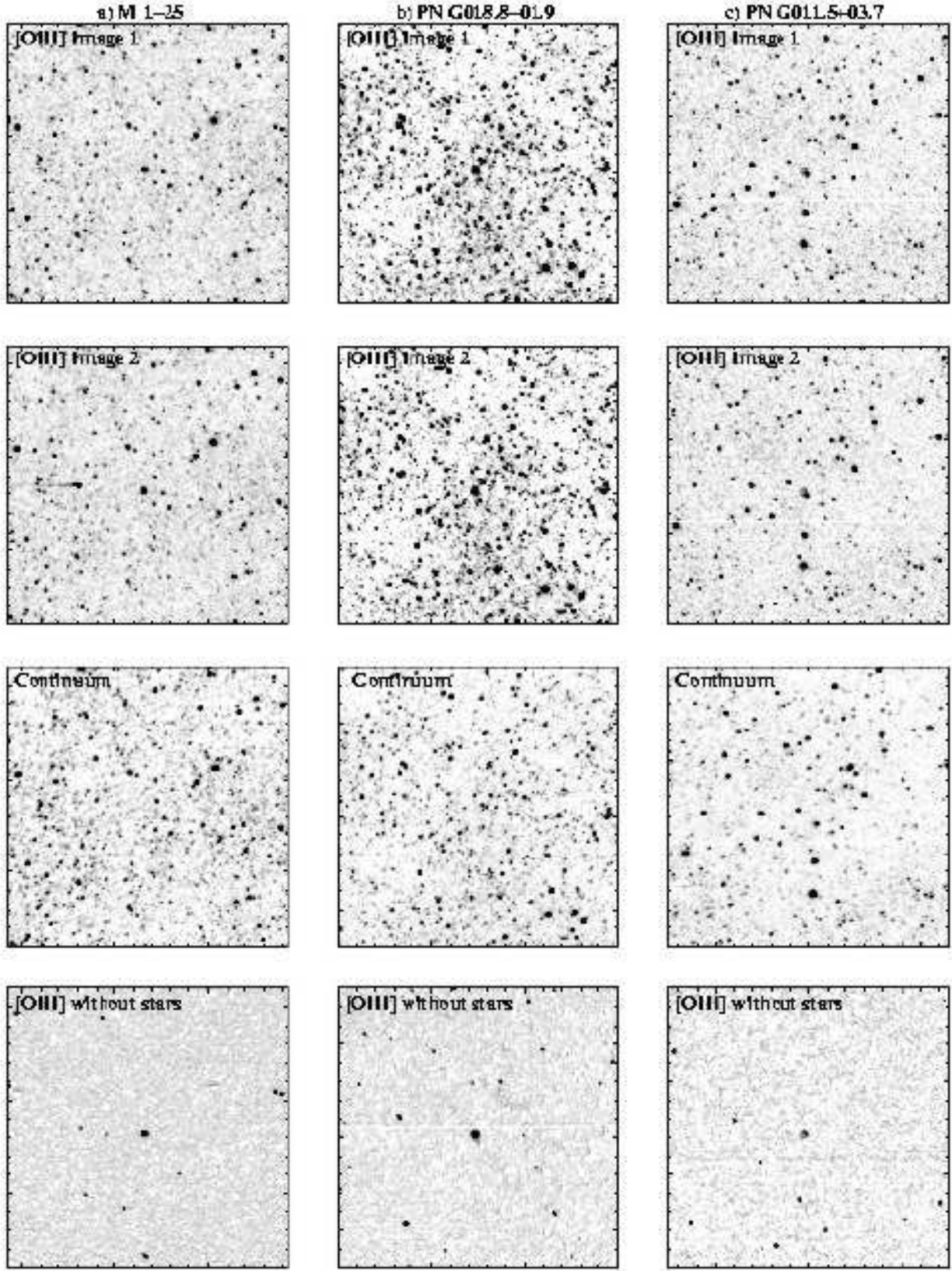}}
\caption[]{Examples of images (20\arcmin\ on both sides each) for three
different objects which show our survey's typical discoveries. The
first set of images (a) is a known PN which was found and presented
for comparison reasons; the other two sets (b) and (c) are new
discoveries. Note that the remaining black dots in the last image of
each set (except the PNe) are cosmic rays which did not remove with
the continuum subtraction. North is at the top, East to the left.} 
\label{fig02}
\end{figure*}

\begin{figure*}
\centering
\mbox{\epsfxsize=6in\epsfbox[0 0 520 429]{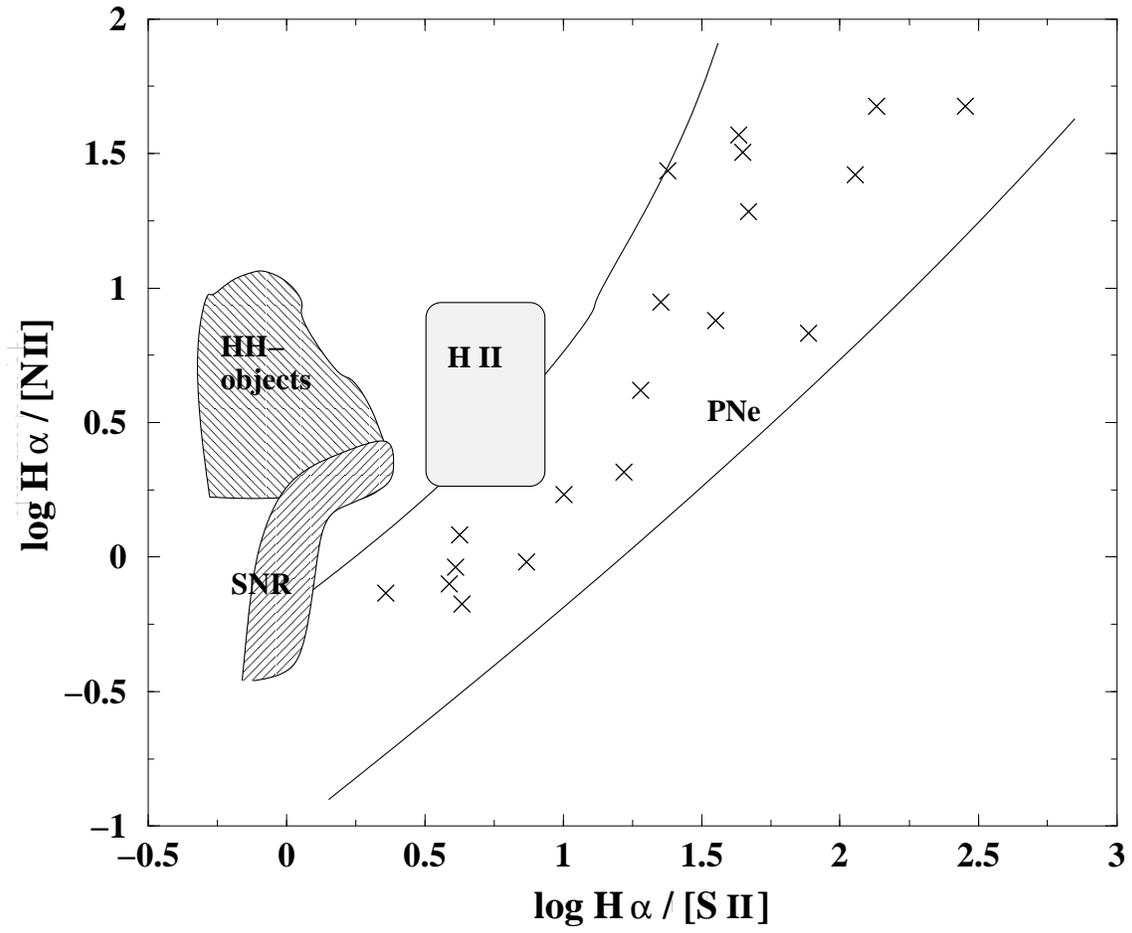}}
\caption[]{Diagnostic diagram (Garcia et al. 1991), where the
positions of the new PNe are shown with a cross (X).} 
\label{fig03}
\end{figure*}

\begin{figure*}
\centering
\mbox{\epsfxsize=6.8in\epsfbox[20 20 575 764]{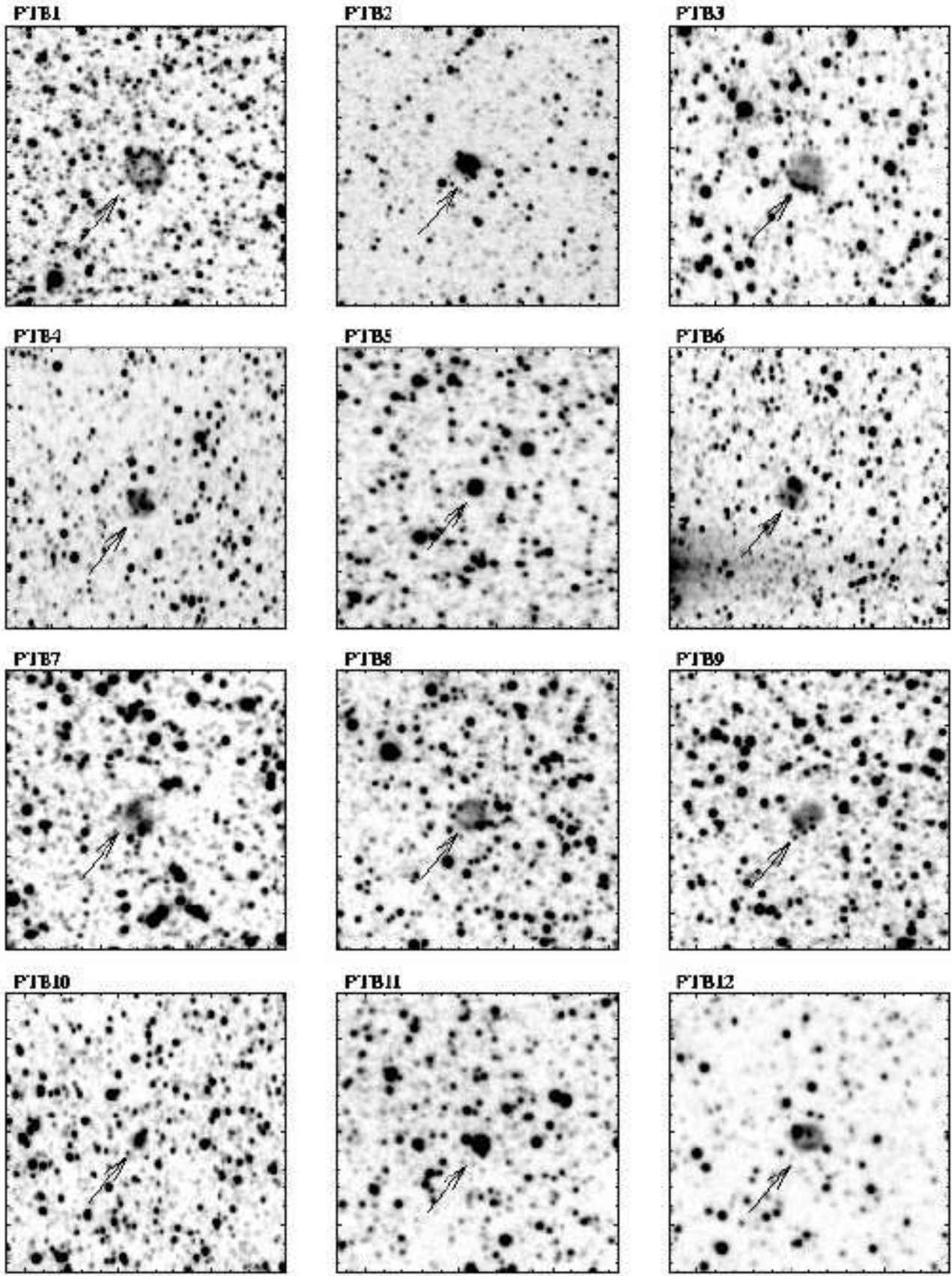}}
\caption[]{\ha $+$\nitrogen~images of all the new PNe taken with the 1.3
m telescope. Black arrows indicate their position at the centre of
each image. The latter, have a size of 150\arcsec on both sides. North
is at the top, East to the left.} 
\label{fig04a}
\end{figure*}

\begin{figure*}
\centering
\mbox{\epsfxsize=6.8in\epsfbox[20 20 575 755]{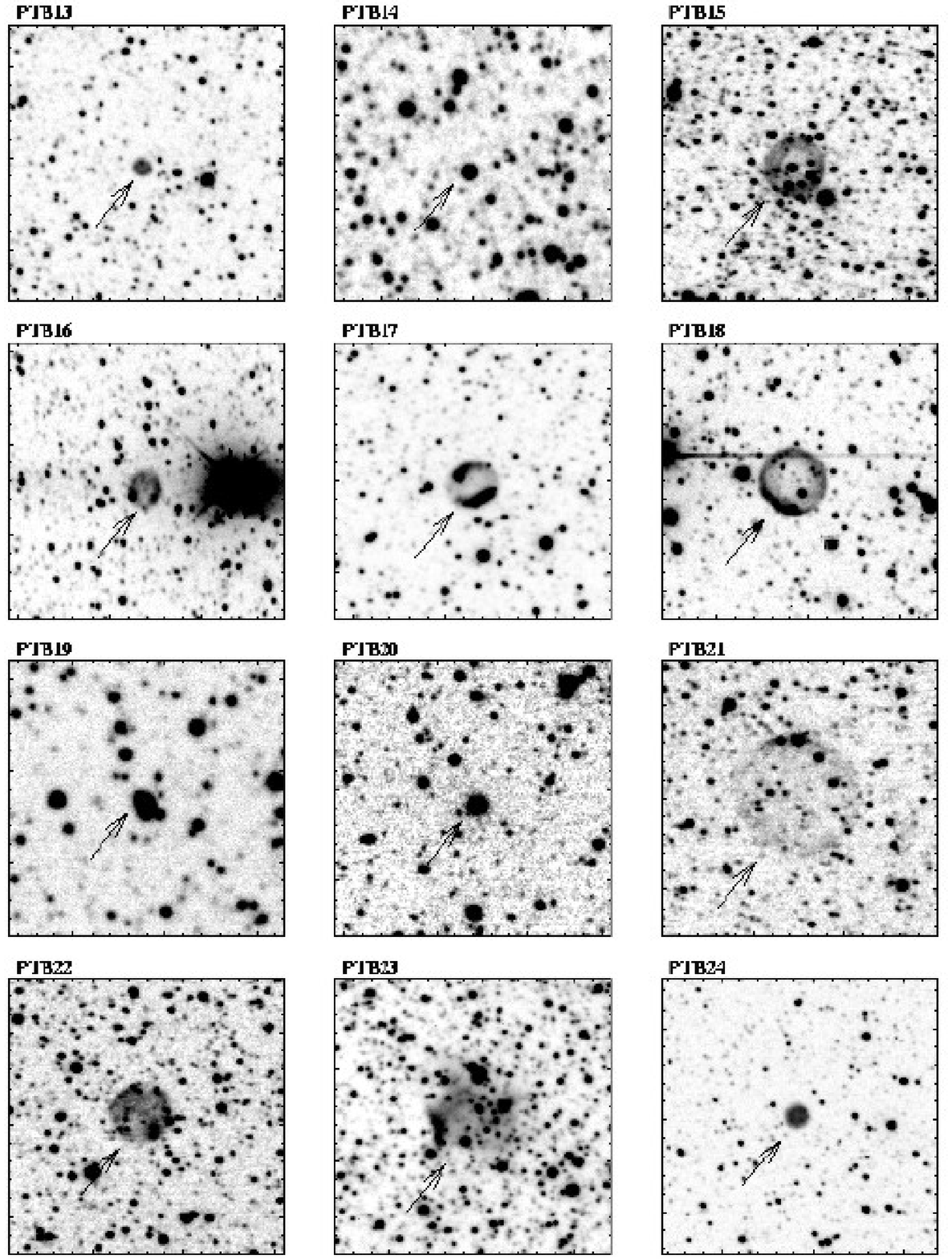}}
\caption[]{Fig. 4 (Continued)} 
\label{fig04b}
\end{figure*}

\begin{figure*}
\centering
\mbox{\epsfxsize=2.4in\epsfbox[0 60 192 279]{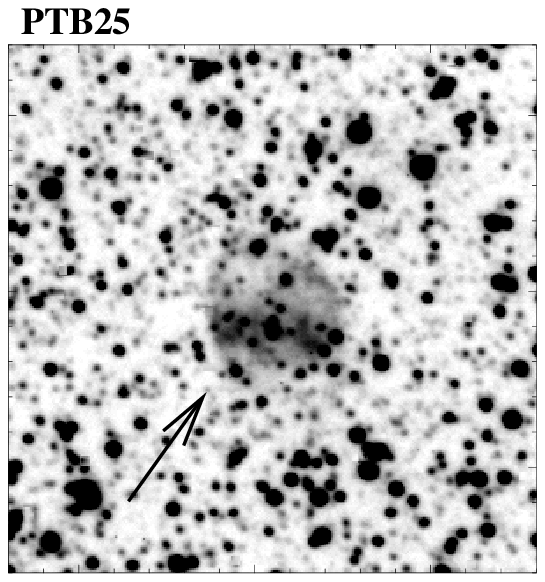}}
\caption[]{Fig. 4 (Continued)} 
\label{fig04c}
\end{figure*}

\begin{figure*}
\centering
\mbox{\epsfxsize=7in\epsfbox[2 96 590 783]{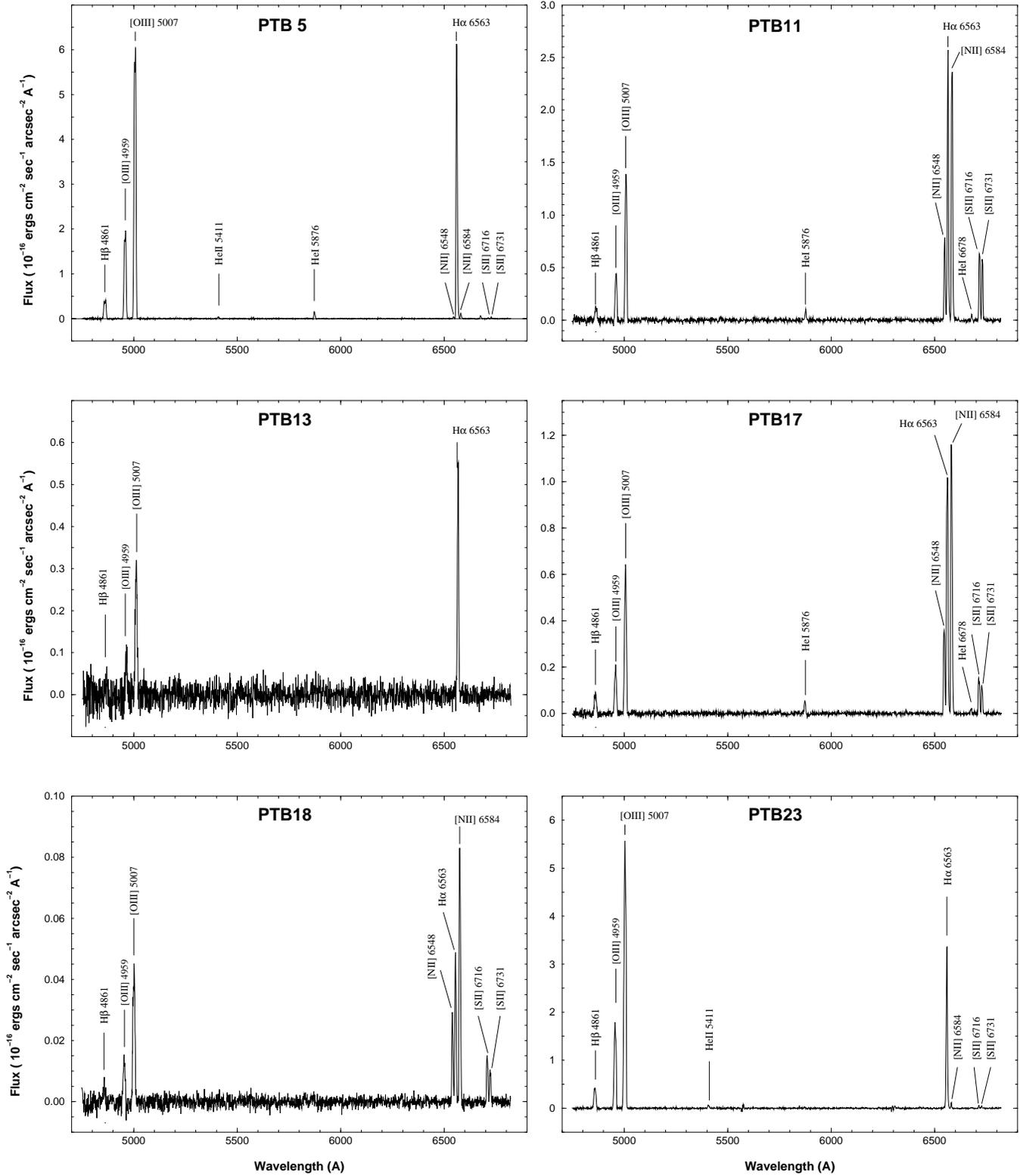}}
\caption[]{Typical observed spectra of our new PNe taken with the 1.3
m telescope. They cover the range of 4750\AA\ to 6815\AA\ and the
emission line fluxes (in units of $10^{-16}$ erg s$^{-1}$ cm$^{-2}$
arcsec$^{-2}$ \AA$^{-1}$) are corrected for atmospheric
extinction. Line fluxes corrected for interstellar extinction are
given in Table 5.}
\label{spectra}
\end{figure*}

\end{document}